\documentstyle[preprint,prl,aps]{revtex}

\begin{document}
\draft
\begin{titlepage}
\preprint{\vbox{\hbox{UDHEP-07-97}
\hbox{July 1997}}}
\title { \large \bf Neutrino Oscillations from String Theory}
\author{\bf A. Halprin$^{(a)}$ and C. N. Leung$^{(a),(b)}$} 
\address{(a) Department of Physics and Astronomy, 
University of Delaware\\
Newark, DE 19716 \\}
\address{(b) Institute of Physics, Academia Sinica, Taipei, 
Taiwan\\} 
\maketitle
\bigskip
\begin{abstract} 
 
We derive the character of neutrino oscillations that results from a model 
of equivalence principle violation suggested recently by Damour and Polyakov 
as a plausible consequence of string theory.  In this model neutrino 
oscillations will take place through interaction with a long range scalar 
field of gravitational origin even if the neutrinos are degenerate in mass.  
The energy dependence of the oscillation length is identical to that in 
the conventional mass mixing mechanism.  This possibility further highlights 
the independence of and need for more exacting direct neutrino mass 
measurements together with a next generation of neutrinoless double beta 
decay experiments.

\end{abstract}
\end{titlepage}

\newpage

Damour and Polyakov \cite{DP} have argued that string theory may very well 
lead to a violation of the equivalence principle through interactions of 
the string dilaton field which may be massless.  They have shown that the 
resultant effective theory of gravity is a variant of Brans-Dicke \cite{BD}
scalar-tensor type and leads to the following two-particle static 
gravitational potential energy,

\begin{equation}
V(r)= - G_Nm_A m_B(1 + \alpha_A\alpha_B)/r,
\end{equation}
where $G_N$ is Newton's gravitational constant.  For vanishing $\alpha_j$ 
the interaction energy is the usual universal spin-2 exchange contribution, 
while the $\alpha$ dependent piece arises from spin-0 exchange.  
The remarkable features of the Damour-Polyakov (DP) scenario are that the 
spin-0 field remains massless and that the $\alpha_j$ are species dependent.  
It is the species dependence that violates the equivalence principle.

An interaction lagrangian density that gives rise to this spin-0 exchange 
contribution to the static gravitational energy is, of course, given by

\begin{equation}
{\cal L} = m_j \alpha_j \overline{\psi_j} \psi_j \phi,
\end{equation}
where $\psi_j$ is a matter field of type $j$.  The dilaton field, $\phi$, 
is coupled to the trace of the energy-momentum tensor through   

\begin{equation}
g_{\mu\nu}\partial^\mu\partial^\nu\phi=-4\pi G_N\alpha_lg_{\mu\nu}T^{\mu\nu}_l,
\end{equation}
where $T^{\mu \nu}_l$ stands for the energy-momentum tensor for $l$ type 
matter.  This interaction embraces the DP scenario up to higher derivatives 
of the gravitational fields and is sufficient for our purposes.

If gravity is treated classically (i.e., as a static background) and 
linearized (i.e., treated as a weak perturbation), the evolution 
of a fermion in an external gravitational field will be governed by a 
Dirac equation with an effective mass $m^*$ given by
 
\begin{equation}
m^* = m - m \alpha \phi_c.
\end{equation}
The classical value of the dilaton field, $\phi_c$, is characterized by 
the $\alpha$ value of the bulk matter producing it and, for a static matter 
distribution, is proportional to the Newtonian potential, $\Phi_N$, viz.

\begin{equation}
\phi_c = \alpha_{\rm ext} \Phi_N.
\end{equation} 
There is also a modification to the metric in the Dirac equation due to 
the spin-2 gravitational field.  But unlike recent alternative 
considerations \cite{G,HL,PHL,IMY,MN,BKL,HLP,MM,MaS,MSa}, the tensorial 
gravity is universal in the DP approach.  It will therefore play no role 
in the neutrino mixing phenomenology of interest here and so we dispense 
with it.

The interaction above can easily be applied to the case of two neutrino 
mixing (e.g., $\nu_e$ and $\nu_\mu$) by replacing $m$ and $m \alpha$ by 
2$\times$2 matrices, $M$ and $M_\alpha$, respectively.  
Let us call the eigenstates of $M$ the mass eigenstates and those of 
$M_\alpha$ the gravitational eigenstates.  To illustrate the possible 
outcome of the DP scenario in neutrino physics, we consider here the 
special case in which the mass and gravitational eigenstates are 
identical and shall be referred to as the $m^*$-eigenstates.  (The more 
general case in which the mass eigenstates are distinct from the 
gravitational eigenstates is similar to the situation discussed in 
Section II.C of Ref.\cite{HLP} and will be dealt with elsewhere.)  
Neutrino flavor oscillatons will therefore take place if the 
$m^*$-eigenstates differ from the neutrino flavor eigenstates and if the 
$m^*$-eigenvalues are not degenerate.  In this case the evolution 
equations governing the oscillation phenomenon of relativistic neutrinos 
are given in the flavor basis by

\begin{equation}
i \frac{d}{dt} 
\left( \begin{array}{c} \nu_e \\ \nu_\mu \end{array} \right)
= {\Delta m^{*2} \over 4E}
\left[ \begin{array}{cc}
- \cos (2\theta) & \sin (2\theta) \\
\sin (2\theta) & \cos (2\theta) \end{array}
\right] 
\left( \begin{array}{c} \nu_e \\ \nu_\mu \end{array} \right),
\end{equation} 
where $\theta$ is a mixing angle and 

\begin{eqnarray}
\Delta m^{*2} 
& \equiv & 
m_2^2 (1 - \alpha_2 \phi_c)^2 - m_1^2 (1 - \alpha_1 \phi_c)^2 \nonumber \\ 
& \simeq & 
\Delta m^2 - 2 \phi_c (m_2^2 \alpha_2 - m_1^2 \alpha_1).
\end{eqnarray}
Here $\Delta m^2 \equiv m_2^2 - m_1^2$ denotes the difference in neutrino 
vacuum masses (squared) and only terms up to first order in $\phi_c$ are 
kept in the above approximation for $\Delta m^{*2}$.  If the vacuum mass 
squared difference dominates $\Delta m^{*2}$, a violation of the 
equivalence principle (VEP) will not be observed in neutrino oscillations.  
On the other hand, even if the neutrinos are completely degenerate but not 
massless, the VEP term will still produce oscillations, and in that case

\begin{equation}
\Delta m^{*2} \simeq -~2 m^2 \alpha_{\rm ext} \Phi_N \Delta \alpha,
\end{equation}
where $m$ is the degenerate neutrino mass and $\Delta \alpha \equiv 
\alpha_2 - \alpha_1$ is the difference between the $\alpha$ values of the 
two neutrino species.

While the physics of the usual spin-2 gravitational field is not dependent 
upon the absolute value of $\Phi_N$, we see that the same is not true for 
the scalar contribution.  Anywhere in our solar system, the dominant contribution to the local gravitational potential appears to come from 
the great attractor which is about $3 \times 10^{-5}$ \cite{MTW,K}.   
For earthbound experiments, we can regard $\Phi_N$ as essentially 
constant.  In this case, the survival probability for an electron 
neutrino that has travelled a distance $L$ is given by 

\begin{equation}
P(\nu_e \rightarrow \nu_e) = 1 - \sin^2 (2\theta) \sin^2(\frac
{L \Delta m^{*2}}{4E}).
\end{equation}
This is analogous to the situation when flavor oscillations of neutrinos 
are caused by their vacuum mass differences.  Consequently, $\Delta m^{*2}$ 
is subject to the same constraints derived for $\Delta m^2$ in the mass 
mixing mechanism.  For instance, according to the analyses in Ref.\cite{BK}, 
the solar neutrino data constrain $\Delta m^{*2}$ to be in the range 

\begin{equation}
4 \times 10^{-6} {\rm eV}^2 < |\Delta m^{*2}| < 10^{-4} {\rm eV}^2
\label{MSWlimit}
\end{equation}
if the MSW transitions \cite{W,MS} are assumed; and in the range 

\begin{equation}
5 \times 10^{-11} {\rm eV}^2 < |\Delta m^{*2}| < 10^{-10} {\rm eV}^2
\label{vaclimit}
\end{equation}
if one assumes vacuum transitions.

To see if the dilaton-induced VEP can have anything to do with the solar 
neutrino deficit, we use the Newtonian potential due to the great attractor 
and the limit on $\alpha_{\rm ext}$ coming from solar-system gravitational 
experiments \cite{R,F}, $\alpha_{\rm ext}^2 < 10^{-3}$.  Since electron 
neutrinos are necessarily involved here, we use the 10 eV limit on its 
mass \cite{PR} as the degenerate value.  These numbers illustrate that 
the VEP mechanism considered here is unlikely to contribute to the solar 
neutrino deficit if the flavor transitions are through the MSW effect.  
On the other hand, if the transitions occur in vacuo, current solar 
neutrino data probe the string theory violation of the equivalence 
principle at the level of 

\begin{equation}
2 \times 10^{-7}~<~|\Delta \alpha|~<~5 \times 10^{-5}.
\label{limit}
\end{equation}
Although not as good as the most restrictive limit for $ordinary\enskip 
matter$ \cite{HEK,AdelS} (which may have little to do with neutrinos), 
this is better than the limits on neutrinos obtained from SN1987A 
\cite{SN87}.  The above limit should not be compared with those obtained 
in Refs. \cite{BKL} and \cite{HLP}, since the equivalence principle 
violation considered there arises from the tensorial gravitational 
couplings whereas the source of VEP here resides in the couplings to the 
string dilaton.

Turning to other neutrino processes, we note that we have considered the 
DP scenario only for the case in which the neutrino vacuum mass terms are 
of the Dirac type.  It follows that the dilatonic contribution to the 
effective mass is also of the Dirac type, so that it will not generate 
neutrinoless double beta decays \cite{HMPR,HPR,HE}.  In the more general 
case in which the vacuum mass terms are of the Majorana plus Dirac type, 
neutrinoless double beta decays as well as neutrino-antineutrino 
oscillations \cite{BP} become possible.  We plan to study the richer 
phenomenology for this case in a future communication.  Finally, it 
should be noted that beta-decay spectrum end-point measurements will see 
the mass $m^*$, which in this scenario could be of order eV rather than 
the scale of $\Delta m^*$.

We therefore conclude there is the distinct possibility that the solar 
neutrino deficit may be telling us about a nonuniversal scalar gravitational 
interaction rather than the existence of a neutrino mass difference.  This 
oscillation mechanism is phenomenologically distinguished from the 
conventional mass mixing mechanism by providing a rationale for the 
possibility that effective neutrino mass differences pertinent to solar 
neutrinos are small while true neutrino masses are orders of magnitude 
larger - with degeneracy protected by a family symmetry.  This adds yet 
another contender to the list of alternatives \cite{CG,GHKLP} to a 
neutrino mass difference and emphasizes further the independence of and 
need for more exacting direct measurements of neutrino masses as well 
as the effective mass arising in neutrinoless double beta decays.  Our 
study here shows that these neutrino experiments are important not only 
for studying the physical properties of neutrinos but also as a means to 
test the low energy phenomenology of string theory.

\vspace*{2.0 cm}
\centerline{\bf Acknowledgement}
\bigskip
This work was supported in part by the U.S. Department of Energy under grant 
DE-FG02-84ER40163.  Part of this work was carried out when CNL was visiting 
SISSA.  He thanks the faculty there, especially S. T. Petcov (and his 
family), for their hospitality.  CNL also wishes to thank M. Fabbrichesi 
and S. T. Petcov for enlightening discussions.


\begin{references}
\bibitem{F}
A much more restrictive bound has been obtained by Y. Fujii, Mod. Phys. 
Lett.  {\bf A 39}, 3685 (1994), but because of the model dependence in 
the calculation we do not use this interesting result here.
\end{references}
\end{document}